\documentstyle[aps]{revtex}

\begin{document}
\draft
\title{The Question of Low-Lying Intruder States in $^8Be$ 
and Neighboring Nuclei}
\author{M.S. Fayache$^{1,2}$, E. Moya de Guerra$^3$, 
P. Sarriguren$^3$, Y.Y. Sharon$^2$, and L. Zamick$^{2,3}$}
\address{(1) D\'{e}partement de Physique, Facult\'{e} des 
Sciences de Tunis\\
Tunis 1060, Tunisia\\
\noindent (2) Department of Physics and Astronomy, Rutgers 
University\\
Piscataway, New Jersey 08855\\
\noindent (3) Instituto de Estructura de la Materia, Consejo 
Superior de\\
Investigaciones Cient\'ificas,\\
Serrano 119, 28006 Madrid, Spain}
\date{\today }
\maketitle

\begin{abstract}
The presence of not yet detected intruder states in $^{8}Be$ 
e.g. a $J=2^{+}$ intruder at 9 $MeV$ excitation would affect 
the shape of the $\beta ^{\mp }$-delayed alpha spectra of 
$^{8}Li$ and $^{8}B$. In order to test the plausibility of 
this assumption, shell model calculations with up to 
$4\hbar \omega $ excitations in $^{8}Be$ (and up to 
$2\hbar \omega $ excitations in $^{10}Be$) were performed. 
With the above restrictions on the model spaces, the 
calculations did not yield any low-lying intruder state in 
$^{8}Be$. Another approach -the simple deformed oscillator 
model with self-consistent frequencies and volume conservation 
gives an intruder state in $^{8}Be$ which is lower in energy 
than the above shell model results, but its energy is still 
considerably higher than 9 $MeV$.
\end{abstract}


\section{Introduction and Motivation}

In an $R$ matrix analysis of the $\beta^{\mp}$-delayed alpha spectra from
the decay of $^8Li$ and $^8B$ as measured by Wilkinson and Alburger \cite
{adw}, Warburton \cite{war} made the following statement in the abstract:
``It is found that satisfactory fits are obtained without introducing
intruder states below 26-$MeV$ excitations''. However, Barker has questioned
this \cite{bar1,bar2} by looking at the systematics of intruder states in
neighboring nuclei. He noted that the excitation energies of $0_2^+$ states
in $^{16}O$, $^{12}C$ and $^{10}Be$ were respectively 6.05 $MeV$, 7.65 $MeV$
and 6.18 $MeV$. Why should there not then be an intruder state in $^8Be$
around that energy?

In recent works \cite{fay1,fay2} the current authors and S. S. Sharma
allowed up to $2 \hbar \omega$ excitations in $^8Be$ and in $^{10}Be$, and
indeed $2p-2h$ intruder states were studied with some care in $^{10}Be$.
Using a simple quadrupole-quadrupole interaction $-\chi Q\cdot Q$ with 
$\chi$=0.3615 $MeV/fm^4$ for $^{10}Be$ and 
$\hbar \omega=45/A^{1/3}-25/A^{2/3}$.
We found a $J=0^+$ intruder state at 9.7 $MeV$ excitation energy. This is
higher than the experimental value of 6.18 $MeV$, but it is in the ballpark.
However, there are other $J=0^+$ excited states below the intruder state
found in the calculation.

In a $0p$-shell calculation with the interaction $-\chi Q\cdot Q$, using a
combination of the Wigner Supermultiplet theory \cite{wig} characterized by
the quantum numbers [$f_1 f_2 f_3$] and Elliott's $SU(3)$ formula \cite
{Elliott}, one can obtain the following expression giving the energies of
the various states:

\begin{equation}
E(\lambda ~\mu )=\bar \chi \left[ -4(\lambda ^2+\mu ^2+\lambda \mu
+3(\lambda +\mu ))+3L(L+1)\right]
\end{equation}
where 
\begin{equation}
\lambda =f_1-f_2,~~~\mu =f_2-f_3
\end{equation}
and 
\begin{equation}
\bar \chi =\chi \frac{5b^4}{32\pi }~~~~(b^2=\frac \hbar {m\omega })
\end{equation}

The two $J=0^+$ states lying below the calculated intruder state in 
$^{10}Be$, at least in the calculation, correspond to two degenerate 
configurations
[411] and [330]. Both of these have configurations $L=1~S=1$ from which one
can form the triplet configurations $J=0^+,~1^+,~2^+$. Hence, besides the
intruder state, we have the above two $J=0^+$ states as candidates for the
experimental $0_2^+$ state at 6.18 $MeV$.

As noted in the previous work \cite{fay1} if, in the $0p$-shell model space
we fit $\chi$ to get the energy of the lowest $2^+$ state in $^{10}Be$ to be
at the experimental value of 3.368 $MeV$ (18$\bar{\chi}$), then the two sets
of triplets are at an excitation energy of 30 $\bar{\chi}$ which equals 5.61 
$MeV$ -not far from the experimental value. There is however a problem -in a 
$0p$-space calculation with $Q \cdot Q$, the lowest $2^+$ state is two-fold
degenerate, corresponding to $J=2^+~K=0$ and $J=2^+~K=2$.

So it is by no means clear if the $0^+$ state in $^{10}Be$ at 6.18 $MeV$ is
an intruder state. We will discuss this more in a later section. It should
be noted that in the previously mentioned calculation \cite{fay2}, the
energy of the intruder state is very sensitive to the value of $\chi$, the
strength of the $Q \cdot Q$ interaction. The energy of this intruder state
drops down rapidly and nearly linearly with increasing $\chi$.

Because of uncertainties due to the truncations in the shell model
calculations, an alternate approach is also considered. This is the deformed
oscillator model with volume conservation and self-consistent frequencies.

\section{Results of the Shell Model Diagonalizations}

In Tables I, II and III we give results for the energies of $J=0^{+}$ 
and $2^{+}$ states in $^{8}Be$, in which up to $4\hbar \omega $
excitations are
allowed relative to the basic configurations $(0s)^{4}(0p)^{4}$. The
different tables correspond to different interactions as follows: \newline
(a) Quadrupole-Quadrupole: $V=-\chi Q\cdot Q$ with 
$\chi =0.3467~MeV/fm^{4}$. \newline
(b) $V=-\chi Q\cdot Q~+~xV_{s.o.}$ ($\chi $ as above and $x=1$). \newline
(c) $V=V_{c}+xV_{s.o.}+yV_{t}$ ($x=1,~y=1$). \newline

Case (c) above consists of a simplified realistic interaction constructed by
Zheng and Zamick \cite{ann}. They took a combination of a central $V_{c}$, a
spin-orbit $V_{s.o.}$ and a tensor interaction $V_{t}$ and fitted the
parameters to the realistic Bonn A bare $G$ matrix elements \cite{bonn}. To
study the effects of varying the spin-orbit and tensor interactions they
multiplied these by factors $x$ and $y$, respectively. For $x=1,~y=1$ one
gets the best fit to the Bonn A matrix elements and this choice is used in
this work. This has been discussed extensively in previous references \cite
{fay1,ann,fay97}.

It should be noted that in all our shell model matrix diagonalizations the
effects of spurious center of mass motion are removed. In the {\sc oxbash}
program used here \cite{oxbash}, this is done by using the Gloeckner-Lawson
method which pushes the spurious states to a very high energy. For more
details see Refs. \cite{fay97,gloek}.

In Tables IV, V and VI we present results for isospin one $J=0^+$ and $2^+$
states in $^{10}Be$ in which up to $2 \hbar \omega$ excitations have been
included. We have the same three interactions as above but with 
$\chi=0.3615~MeV/fm^4$ in (a) and (b).

In all the tables we give the excitation energies of the $J=0^+$ and $2^+$
states and the percent probability that there are no excitations beyond the
basic configuration ($0 \hbar \omega$) and the percentage of $2 \hbar \omega$
excitations (as well as $4 \hbar \omega$ excitations for $^8Be$).

Note that for interaction (a) the respective percentages for the ground
state of $^8Be$ (see Table I) are 62.8\%, 25.7\% and 11.5\%: there is
considerable mixing. Thus we should not forget, when we discuss the question
``where are the intruder states?'', that there is considerable admixing of
$2 \hbar \omega$ and $4 \hbar \omega$ excitations {\em in the ground state}.
Note that the ground state configuration does not change very much for the
three interactions that are considered here. For example, as seen in Table
III, the corresponding percentages for the ($x,y$) interaction are 62.2\%,
26.2\% and 11.6\%.

By looking at these tables, it is not too difficult to see at what energies
the intruder states set in. One sees a sharp drop in the $0\hbar \omega $
occupancy. For example in Table I, whereas the $0\hbar \omega $ percentage
for the 18.7 $MeV$ and 20.2 $MeV$ states are respectively 93.9\% and 94.6\%,
for the next state at 26.5 $MeV$ the percentage drops to 29.4\% -also the
next four states listed have very low $0\hbar \omega $ percentages and are
therefore intruders.

The terminology {\em intruder state} is somewhat arbitrary. It is used by
experimentalists to refer to certain low-lying states with certain
properties. In shell model calculations it is generally used for states
whose main components are outside the model space composed of one major
shell $N$ (the valence shell). Following this criterion in our theoretical
calculations we define an intruder state as one for which the $0\hbar \omega 
$ percentage is less than 50\%. By this criterion, and for the three
interactions discussed here, the lowest $J=0^{+}$ intruder states in $^{8}Be$
are at 26.23 $MeV$, 26.5 $MeV$ and 28.7 $MeV$ (see Tables I,II, and III).
The $J=2^{+}$ intruder states are at 27.15 $MeV$, 27.5 $MeV$ and 33.7 $MeV$.
Note that up to 4$\hbar \omega $ excitations were allowed in these
calculations. These energies are very high and would argue against the
suggestion by Barker that there are low-lying intruder states in $^{8}Be$.

What about $^{10}Be$? Remember that in this nucleus we only include up to 
$2\hbar \omega $ excitations. For the three interactions considered, the
lowest $J=0^{+}~T=1$ intruder states are at 9.7 $MeV$, 11.4 $MeV$ and 31.0
$MeV$. The `anomalous' behavior for the last value (31.0 $MeV$ for the 
($x,y$) interaction) will be discussed in a later section.

Note that when a spin-orbit is added to $Q \cdot Q$, the energy of the
intruder state goes up $e.g.$ 11.4 $MeV$ $vs$ 9.7 $MeV$. The lowest-lying 
$J=2^+~T=1$ intruder states are at 11.9 $MeV$, 13.8 $MeV$ and 33.4 $MeV$. The
energy of the non-intruder ($L=1~S=1$) $J=0^+,~1^+,~2^+$ triplet also goes
up as can be seen from Tables IV and V.

For the two $Q \cdot Q$ interactions, the energies of the intruder states in 
$^{10}Be$ are much lower than in $^8Be$. This conclusion still holds if we
were to use $^8Be$ energies calculated in (0+2)$\hbar \omega$ configuration
space -see Table VII. This would indicate that even if we do find low-lying
intruder states in $^{10}Be$, such a finding in itself is not proof that
they are also present in $^8Be$. Indeed, our calculations would dispute this
claim.

\section{$(0+2) \hbar \omega$ $vs$ $(0+2+4) \hbar \omega$ Calculations for
$^8Be$}

In Table VII we show the results for the energy of the first intruder state
in $^8Be$ in calculations in which only up to $2 \hbar \omega$ excitations
are included and compare them with the corresponding results for up to 
$4\hbar \omega$. For interactions (a) and (b), the value of $\chi$ was
changed to 0.4033 $MeV/fm^4$ in order that the energy of the $2_1^+$ state
come close to experiment. In more detail, we have to rescale $\chi$
depending on the model space in order to get the $2_1^+$ state at the right
energy. In general, the more $np-nh$ configurations we include the smaller
$\chi$ is.

We see that in the larger-space calculation $(0+2+4)\hbar \omega $, the
energies of the lowest intruder states in most cases come down about 5 $MeV$
relative to the $(0+2)\hbar \omega $ calculation. The excitation energies
are still quite high, however, all being above 25 $MeV$. One possible reason
for the difference between the results of the two calculations is that in
the $(0+2)\hbar \omega $ calculation there is level repulsion between the 
$0\hbar \omega $ and the $2\hbar \omega $ configurations, and that the 
$4\hbar \omega $ configurations are needed to push the $2\hbar \omega $
states back down.

\section{The first excited $J=0^+$ state of $^{10}Be$}

Is the first excited $J=0^{+}$ state in $^{10}Be$ an intruder state or is it
dominantly of the $(0s)^{4}(0p)^{6}$ configuration? Experimentally, very few
states have been identified in $^{10}Be$. The known positive-parity states
are as follows \cite{ajz}:

\begin{center}
$
\begin{tabular}{cc}
$J^{\pi }$ & $E_{x}(MeV)$ \\ 
$0_{1}^{+}$ & $0.000$ \\ 
$2_{1}^{+}$ & $3.368$ \\ 
$2_{2}^{+}$ & $5.959$ \\ 
$0^{+}$ & $6.179$ \\ 
$2^{+}$ & $7.542$ \\ 
$\left( 2^{+}\right) $ & $9.400$
\end{tabular}
$
\end{center}

In the $(0s)^4(0p)^6$ calculation with a $Q\cdot Q$ interaction, the
lowest $2^{+}$ state at 18$\bar \chi $ is doubly degenerate and
corresponds to $K=0$
and $K=2$ members of the [42] configuration. There are two degenerate 
($L=1~S=1$) configurations at 30$\bar \chi $ with supermultiplet
configurations [330] and [411]. From $L=1~S=1$ one can form a triplet of
states with $J=0^{+},~1^{+},~2^{+}$. If we choose $\bar \chi $ by getting
the $2_1^{+}$ state correct at 3.368 $MeV$, then the two $L=1~S=1$ triplets
would be at $30/18\times 3.36~MeV=5.61~MeV$. However, there should be a 
{\em triplet} of states. In more detailed calculations, as the spin-orbit
interaction is added to the $Q\cdot Q$ interaction, the triplet degeneracy
gets removed with the ordering $E_{2^{+}}<E_{1^{+}}<E_{0^{+}}$. As seen in
Table IV, the $J=0^{+}$ and $2^{+}$ states of $^{10}Be$ at 3.7 $MeV$ and 7.3 
$MeV$ are degenerate with a pure $Q\cdot Q$ interaction. This is also true
for $J=1^{+}$. In Table V, however, when the spin-orbit interaction is added
to $Q\cdot Q$, we find that whereas the $0_2^{+}$ is at 8.0 $MeV$, the 
$2_3^{+}$ state is at 6.8 $MeV$.

Hence if the $0^+$ state at 6.179 $MeV$ were dominantly an $L=1~S=1$
non-intruder state, one would expect a $J=1^+$ and a $J=2^+$ state at lower
energies. Thus far no $J=1^+$ level has been seen in $^{10}Be$ but this is
undoubtedly due to the lack of experimental research on this target. Now
there {\em is} a lower $2^+$ state at 5.959 $MeV$. This could be a member of
the $L=1~S=1$ triplet {\em or} it could be the $K=2$ state of the [42]
configuration.

Hence, one possible scenario is that indeed the $2^+_2$ state is dominantly
of the [42] configuration and the $J=0_2^+$ state is a singlet. This would
support the idea that the $J=0^+_2$ state is an intruder state. The second
scenario has the $J=2_2^+$ state being dominantly an $L=1~S=1$ state for
which the $J=1^+$ member has somehow not been found. This would be in
support of the idea that the $0_2^+$ state is {\em not} an intruder state.

Let us look in detail at Tables IV, V and VI which show where the energies
of the intruder states are in a (0+2)$\hbar \omega$ calculation. For the 
$Q\cdot Q$ interaction (with $\chi=0.3615~MeV/fm^4$), the lowest $J=0^+$
intruder state is at 9.7 $MeV$ and the lowest $J=2^+$ intruder state is at
11.9 $MeV$. These energies are {\em much lower} than the corresponding
intruder state energies for $^8Be$. This in itself is enough to tell us that
the presence of a low-energy intruder state in $^{10}Be$ does not imply that
there should be a low energy intruder state in $^8Be$. Note that the
intruder states in this model space and with this interaction have 100\% 
`2$\hbar \omega$' configurations. This has been noted and discussed in \cite
{fay2} and is due to the fact that the $Q \cdot Q$ interaction cannot excite
two nucleons from the $N$ shell to the $N\pm 1$ shell.

Still, in Table IV, there are two $J=0^+$ states (below the intruder state)
at 3.7 $MeV$ and 7.3 $MeV$. Even in this large-space calculation, these are
members of degenerate $L=1~S=1$ triplets $J=0^+,~1^+,~2^+$. Indeed, if we
look down the table, we see the 3.7 $MeV$ and 7.3 $MeV$ values in the $J=2^+$
column.

In Table V, when we add the spin-orbit interaction to $Q \cdot Q$, the
energies of the $0^+_2$ and $0^+_3$ states go up, but so does the energy of
the $J=0^+_4$ intruder state. The energies of the $0^+_2$, $0^+_3$ and 
$0^+_4 $ (intruder) states in Table IV are 3.7, 7.3 and 9.7 $MeV$; in Table
V, with the added spin-orbit interaction they are 8.0, 9.6 and 11.4 $MeV$.

In Table VI we show results of an up-to-2$\hbar \omega$ calculation with the
realistic interaction. Here, we see a drastically different behavior for the
intruder state energy in $^{10}Be$. The lowest $J=0^+$ intruder state is at
31.0 $MeV$, and the lowest $J=2^+$ intruder state is at 33.4 $MeV$ (recall
our operational definition -an intruder state has less than 50\% of the 
0$\hbar \omega$ configuration). For the $Q \cdot Q$ interaction, in contrast,
the intruder state was at a much lower energy. A possible explanation is
that for the ($x,y$) interaction, unlike $Q \cdot Q$, one {\em does have}
large off-diagonal matrix elements in which two nucleons are excited from $N$
to $N\pm 1$ $e.g.$ from $0p$ to $1s-0d$. This will cause a large level
repulsion between the 0$\hbar \omega$ and the 2$\hbar \omega$ configurations
and drive them far apart. Presumably, if we included 4$\hbar \omega$
configurations into the model space, they would push the 2$\hbar \omega$
configurations back down to near their unperturbed positions.

Thus, the problem is rather difficult to sort out theoretically, so we can
at best suggest that more experiments be done on $^{10}Be$. For example, the 
$B(E2)$ to the $2_2^+$ state would be useful. There should be a much larger
B(E2) to the $L=2~K=2$ member of a [42] configuration than to an ($L=1~S=1$)
state. We also predict a substantial $B(M1)\uparrow$ to the first $J=1^+~T=1$
state in $^{10}Be$. Whereas with a pure $Q \cdot Q$ interaction the $B(M1)$
to this state would be zero, the presence of a spin-orbit interaction will
`light up' the $1^+_1$ state in $^{10}Be$. The $J=1^+$ should be seen.

\section{The Deformed Oscillator Model with Volume Conservation and
Self-Consistent Frequencies}

As an alternative to the shell model approach for finding the energies of
intruder states, we consider the deformed oscillator model of Bohr and
Mottelson \cite{bm}. The Hamiltonian is a sum of one-body terms, one of
which is

\begin{equation}
H=-\frac{\hbar ^2}{2m}\nabla ^2+\frac m2(\omega _x^2x^2+\omega
_y^2y^2+\omega _z^2z^2)
\end{equation}
Furthermore, we assume volume conservation: 
\begin{equation}
\omega _x\omega _y\omega _z=\omega _0^3\equiv constant
\end{equation}

The intrinsic energy is given by

\begin{equation}
E_{int}=\Sigma _x\hbar \omega _x+\Sigma _y\hbar \omega _y+\Sigma _z\hbar
\omega _z
\end{equation}
where $\Sigma_x=\sum (N_x+1/2)$ where $N_x$ is the number of quanta in the
$x-$direction.

The self-consistency condition is

\begin{equation}
\Sigma _x\omega _x=\Sigma _y\omega _y=\Sigma _z\omega _z
\end{equation}

This can be obtained by minimizing the kinetic energy -indeed for a two-body
delta interaction the potential energy depends only on $\omega _0$ and not
on the deformation. With this condition, the energy is given by 
$E_{int}=3\Sigma _z\hbar \omega _z=3\hbar \omega _0\left( \Sigma _x\Sigma
_y\Sigma _z\right) ^{1/3}$.

For a simple estimate, we take $\hbar \omega _0=45A^{-1/3}-25A^{-2/3}$. This
model has been previously applied by L. Zamick {\em et. al.} \cite{zam}.

The calculations for the intrinsic states are remarkably simple. One just
has to evaluate $\Sigma _x,~\Sigma _y$ and $\Sigma _z$ for the ground state
and the intruder states. The single-particle states are classified as 
$(N_x,N_y,N_z)$. The relevant ones for this calculation are (0,0,0), (0,0,1),
(1,0,0), (0,1,0) and (0,0,2). For example, for the ground state of $^8Be$,
the states (0,0,0) and (0,0,1) are occupied so that one has:

$\Sigma_x=4\times1/2+4\times1/2=4$

$\Sigma_y=4\times1/2+4\times1/2=4$

$\Sigma _{z}=4\times 1/2+4\times 3/2=8$

\noindent For the $2p-2h$ intruder states, there are four nucleons in
(0,0,0), two in (0,0,1) and two in (0,0,2). Hence,

$\Sigma_x=\Sigma_y=8\times1/2=4$

$\Sigma _{z}=4\times 1/2+2\times 3/2+2\times 5/2=10$

\noindent For the ground state, the volume conservation condition ($\omega
_{x}\omega _{y}\omega _{z}=\omega _{0}^{3}$) becomes:

$8/4\times 8/4 \times \omega_z^3=\omega_0^3$ and

\[
\frac{\omega_z}{\omega_0}=0.62996 
\]
The intrinsic state energy is then $E=3\times 8\times 0.62996\hbar \omega
_0=15.1990\hbar \omega _0.$ The calculations for other states and other
nuclei are carried out in the same way.

In order to compare our results with experiment we must obtain the energies
of the $J=0^{+}$ and $J=2^{+}$ states. The $0^{+}$ and $2^{+}$ energies are
computed as follows.

In the {\bf axial case}, for a given intrinsic configuration,

\begin{equation}
E_{2^{+}}-E_{0^{+}}=\frac 3{{\cal I}}\;,
\end{equation}

\begin{equation}
E_{0^{+}}=E_{int}-\Delta E_R
\end{equation}
where the zero point energy \cite{moyazam}

\begin{equation}
\Delta E_R=\frac{\left\langle J^2\right\rangle }{2{\cal I}}
\end{equation}
with $\left\langle J^2\right\rangle $ the expectation value of the angular
momentum squared

\begin{equation}
\left\langle J^2\right\rangle =\left\langle J_{\perp }^2\right\rangle
=\left\langle J_x^2\right\rangle +\left\langle J_y^2\right\rangle
=2\left\langle J_x^2\right\rangle
\end{equation}
and ${\cal I}$ the cranking moment of inertia for the corresponding
configuration, i.e.,

\begin{equation}
\left\langle J_x^2\right\rangle =\sum_{ph}\left| \left\langle p\left|
j_x\right| h\right\rangle \right| ^2
\end{equation}

\begin{equation}
{\cal I}={\cal I}_x=2\sum_{ph}\frac{\left| \left\langle p\left| j_x\right|
h\right\rangle \right| ^2}{\epsilon _p-\epsilon _h}
\end{equation}
with $h$ and $p$ the occupied and unoccupied states, respectively, in the
configuration at hand.

In the {\bf triaxial} {\bf case} (see for instance \cite{greiner}) there are
two $2^{+}$ states

\begin{equation}
E_{2^{+}}-E_{0^{+}}=\left( \frac 1{{\cal I}_x}+\frac 1{{\cal I}_y}+ \frac 
1{{\cal I}_z}\right) \left\{ 1\mp \left[ 1-\frac 38 
\frac{2{\cal I}_x{\cal I}_z{\cal I}_y\left( 4{\cal I}_x+
4{\cal I}_y+ 3{\cal I}_z\right) +{\cal I}_z^2\left( 
{\cal I}_x^2+{\cal I}_y^2\right) } {\left( {\cal I}_x{\cal I}_z+
{\cal I}_y{\cal I}_z+{\cal I}_x {\cal I}_y\right) ^2}\right] \right\} ^{1/2}
\end{equation}
The lowest of these $2^{+}$ states is given in the table for the case of the
triaxial configuration in $^{10}Be$ and can be also obtained from the
simpler equation

\begin{equation}
E_{2^{+}}-E_{0^{+}}\simeq \frac 32\left( \frac 1{{\cal I}_x}+\frac 1 
{{\cal I}_y}\right)
\end{equation}
The zero point energy in the triaxial case is obtained as

\begin{equation}
\Delta E_R=\left( \frac{\left\langle J_x^2\right\rangle }{{\cal I}_x}+ 
\frac{\left\langle J_y^2\right\rangle }{{\cal I}_y}+\frac{\left\langle
J_z^2\right\rangle }{{\cal I}_z}\right)
\end{equation}

\section{Discussion of results}

\subsection{Experimental situation}

We present results for $^8Be,$ $^{10}Be,$ and $^{12}C.$ The latter nucleus
is included because there is a known $J=0^{+}$ intruder state at 7.654 MeV,
generally considered to be a $4p-4h$ state. In $^{10}Be$ there is a $J=0^{+}$
excited state at 6.11 MeV, which may well be a $2p-2h$ intruder state.
However $^{10}Be$ is a remarkably understudied nucleus and it would be nice
to have more experimental work to confirm (or deny) this. Although we will
not include calculations for $^{11}Be$ here, it should be noted that for
this nucleus there is an inversion with a $J=1/2^{+}$ ground state, which is
0.3196 MeV below the {\em expected parity} $J=1/2^{-}$ state. This is
unmistakable evidence that there are low-lying intruders in this region.

\subsection{The calculation}

We present the results for the deformed oscillator model in Table VIII. This
table contains both the input parameters and the results for the intrinsic
state energies, and the energies of the $J=0^{+}$ and $J=2^{+}$ intruder
states.

We first give $\Sigma _{x},$ $\Sigma _{y},$ $\Sigma _{z}$ from which the
frequencies $\omega _{x},$ $\omega _{y},$ $\omega _{z},$ and $\omega _{0}$
are obtained. This is sufficient to obtain the intrinsic state energies in
units of $\hbar \omega _{0}$. Next the quantities needed to get the energies
of the $J=0^{+}$ and $J=2^{+}$ states are shown. These are the expectation
values $\left\langle J_{x}^{2}\right\rangle ,\,\left\langle
J_{y}^{2}\right\rangle $ and $\left\langle J_{z}^{2}\right\rangle $ and the
moment of inertia in units of $\left( \hbar \omega _{0}\right) ^{-1}.$ We
then present the zero point energy $\Delta E_{R}$ in units of $\left( \hbar
\omega _{0}\right) $. We then present $\left( \hbar \omega _{0}\right) $
using the formula $\hbar \omega _{0}=45\,A^{-1/3}-25\,A^{-2/3}\,MeV.$ It
would be better to fit $\hbar \omega _{0}$ to experiment. However, since 
$^{8}Be$ is unstable one cannot measure the r.m.s. radius. There is no data
available for $^{10}Be$ and for $^{12}C$ the error bars on r.m.s. are fairly
large. At any rate, since we next present results for $E_{J=0}^{*}$ both in
units of $\left( \hbar \omega _{0}\right) $ and in MeV, it is easy for the
reader to obtain results for an $\hbar \omega _{0}$ of his/her choice. We
lastly give the excitation energies of the $J=2^{+}$ states.

Let us first discuss $^{12}C$ because the experimental situation here is
most solid. The values of $\Sigma _x,~\Sigma _y$ and $\Sigma _z$ for the
ground state are 10, 6, 10. This implies $\omega _x=\omega _z<\omega _y.$
This means that the $y$-axis is the symmetry axis and the nucleus will be
oblate. The values of $\Sigma _x,~\Sigma _y$ and $\Sigma _z$ for the $4p-4h$
intruder state are 6, 6, 18. Hence the $z$-axis will be the symmetry axis
and the intrinsic state is prolate. We obtain the excitation energy of the
$4p-4h\,\,\;J=0^{+}$ state to be $E_{J=0^{+}}^{*}=6.55\,MeV.$ The
experimental value is 7.65 MeV. Considering the simplicity of this model the
agreement is remarkable, and we must take the predictions of this model
seriously, even if we do not fully understand why it works so well.

Rather than use the approximate formula $\hbar \omega _0=\left(
45\,A^{-1/3}-25\,A^{-2/3}\,\right) MeV$ we can for a given nucleus fit the
mean square charge radius, provided this quantity has been measured. This is
not the case for $^8Be$ (unstable) or $^{10}Be$, but for $^{12}C$ De Vries
et al. \cite{devries} give three results due to different groups, 
$\left\langle r^2\right\rangle =2.472\left( 15\right) ,\;2.471\left( 
6\right) $ and $2.464\left( 12\right) $ fm.

In our formulation the charge radius is given by

\begin{equation}
\left\langle r^2\right\rangle _{ch}=\frac{\hbar ^2}{Zm}\left( 
\frac{\Sigma_{\pi z}}{\hbar \omega _z}+\frac{\Sigma _{\pi x}} {\hbar 
\omega _x} +\frac{\Sigma _{\pi y}}{\hbar \omega _y}\right)
\end{equation}
If we take $\left\langle r^2\right\rangle ^{1/2}=2.47$ fm, we find $\hbar
\omega _0=15.85$ MeV. This is larger than the value in Table VIII. We now
find that the excitation energy of the $J=0^{+}\;4p-4h$ state is 6.97 MeV.
This is closer to the experimental value of 7.654 MeV, than the value using
the approximate formula for $\hbar \omega _0\;(6.55MeV).$

For $^{10}Be$ the values of $\Sigma _x,~\Sigma _y$ and $\Sigma _z$ for the
ground state are 7, 5, 9; for the $2p-2h$ intruder state they are 5, 5, 13.
Thus the ground state band is triaxial but the intruder state has axial
symmetry. We obtain $E_{J=0^{+}}^{*}=6.36\,MeV$ in close agreement with the
experimental result of 6.11 MeV.

We also include results for the axial symmetry approximation for the ground
state of $^{10}Be.$ We replace the numbers 7, 5, 9 by 6, 6, 9. This might
seem like a modest change. However, this is not the case. Indeed we find
that the $2p-2h$ intruder state is 4.03 MeV below the axial ground state.
This is due to a combination of reasons. First, the axial intrinsic ground
state is 3.1 MeV above the triaxial intrinsic ground state. Secondly, we get
a large zero point shift in the triaxial case because we get contributions
from all three axes in the expression $\Delta E_{R}=\Delta E_{x}+\Delta
E_{y}+\Delta E_{z}.$ Again, if we had made the axial approximation for the
$0p-0h$ state we would have reached the erroneous conclusion that the $2p-2h$
intruder state was the ground state. By correctly taking into account the
triaxiality the situation gets reversed.

We now come to our main focus, the intruder states in $\ ^8Be.$ We consider
both the $2p-2h$ and the $4p-4h$ intruders. We find that the excitation
energies are much higher than in $^{10}Be$ or $^{12}C.$ The $J=0^{+}$ $2p-2h$
state is at 17.23 MeV and the $J=0^{+\text{ }}4p-4h$ state is at 32.34 MeV
in this calculation. We can understand this behaviour by considering the
Nilsson diagram shown in Fig. 1. For $^{10}Be$ and $^{12}C$ we take nucleons
from upward-going lines in the $p$ shell and put them into a down-going line
in the $s-d$ shell. The energy required to do this is much less for finite
$\beta $ than it is for $\beta =0$, as can be easily seen from Fig. 1. For
$^8Be,$ on the other hand we must take 2 nucleons from a down-going Nilsson
line. This obviously costs much more energy. The figure and the
corresponding argument make it quite convincing that the presence of
low-lying intruder states in $^{10}Be$ and $^{12}C$ does not imply that
there will be low-lying intruders in $^8Be.$

\section{Conclusions}

Because of the important implications to astrophysics of the $^8Be$ nucleus,
we feel that Barker's suggestion \cite{bar1,bar2} to worry about the
presence of low-lying intruder states in this and neighboring nuclei is
quite sensible. However, our calculations do not support the presence of
low-lying intruder states in $^8Be,$i.e., of a $J=2^{+}$ intruder at 9 MeV
(which would also imply a $J=0^{+}$ intruder at 6 MeV). Our lowest $J=0^{+}$
intruder in the deformed oscillator model is above 17 MeV and the $J=2^{+}$
above 19 MeV. These energies are lower than the 26 MeV gate mentioned by
Warburton in the abstract of his 1986 work \cite{war}, but are sufficiently
high so as not to seriously affect the alpha spectrum.

Our case was made more convincingly the fact that the same calculation does
yield low-lying intruders in $^{10}Be$ and $^{12}C.$ In $^{12}C$ we are in
close agreement with experiment (6.55 MeV vs. 7.654 MeV exp.). In $^{10}Be$
our calculated $J=0^{+\text{ }}$state energy is very close to that of the
first excited $0^{+}$ state (6.36 MeV vs. 6.111 MeV exp.). However more
experimental work will have to be done to determine if this is indeed an
intruder state. Another possibility is that the 6.11 MeV state is the 
$J=0^{+}$ member of an $L=1,S=1$ triplet with orbital symmetry 
$\left[411\right] $ or $\left[ 331\right] $.

Some questions remain. Why are the shell model energies higher than the
deformed oscillator ones. It may be due to the truncated space used in the
shell model calculations. If this is the case then this indicates a rather
slow convergence. It would be of interest to try to enlarge the model space
to test out this idea. It should be emphasized that in the $Q\cdot Q$
calculations the parameter $\chi $ was chosen carefully so that the energy
of the first $2^{+}$ state came out correctly. As we enlarge the model space
we choose $\chi $ so that the fit to the $2^{+}$state is maintained. This
means that $\chi $ becomes smaller as the model space is increased.

We lastly express wonderment that the deformed oscillator model, with zero
point energy corrections, seems to work so well in getting the intruder
states at close to the right energies. In shell model calculations with
realistic interactions it is very difficult to get the intruder states to
come down low enough. This is because one starts with a spherical basis
where for say $^{12}C$ the starting point energy for the $4p-4h$ state is 
$4\hbar \omega =59.5MeV.$ One has to get the state down to 7.65 MeV and this
is very difficult. It would be interesting to see whether this can be done
with other realistic interactions suitably tailored for these type of
calculations. In any case, the model space to do this must be enormous.
However the deformed oscillator model almost effortlessly gets the state
close to this energy. The Nilsson diagram in Fig. 1 explains in part this
success but it would be nice to have a more quantitative understanding.

\section{Acknowledgements}

We are extremely grateful to Ian Towner for suggesting this problem. This
work was supported by a Department of Energy Grant No. DE-FG02-95ER40940, by
DGICYT (Spain) grants under contracts Nos PB95/0123 and SA95-0371 and by a
Stockton College Distinguished Faculty research grant. M.S. Fayache kindly
acknowledges travel support from the Laboratoire de la Physique de la
Mati\`{e}re Condens\'{e}e at the Universit\'{e} de Tunis, Tunisia.

\nopagebreak

\pagebreak

\begin{center}
{\bf Figure Captions}
\end{center}

{\bf Figure 1.-} Schematic Nilsson energies as a function of deformation.

\pagebreak

\begin{table}

{\bf Table I. }$J=0^{+}$ and $2^{+}$ states in $^8Be$ for the interaction 
$-\chi Q\cdot Q$ with $\chi =0.3365~MeV/fm^4$ with up to $4\hbar \omega $
excitations allowed. The percentage of $0\hbar \omega $, $2\hbar \omega $
and $4\hbar \omega $ occupancies are given, as well as the 
$B(E2)(0_1^{+}\rightarrow 2_i^{+}$).

\begin{center}
\begin{tabular}{ccccc}
\multicolumn{5}{c}{(a) $J=0^{+}~T=0$ States} \\ 
{$E_{exc}(MeV)$} & $0~\hbar \omega $ & $2~\hbar \omega $ & $4~\hbar \omega $
&  \\ \hline
0.00 & 64.6 & 24.6 & 10.7 &  \\ 
11.37 & 83.4 & 10.9 & 5.7 &  \\ 
15.88 & 94.4 & 2.1 & 3.5 &  \\ 
17.86 & 94.3 & 2.5 & 3.2 &  \\ 
19.38 & 94.9 & 2.1 & 3.0 &  \\ 
26.23 & 28.5 & 50.9 & 20.6 &  \\ 
29.70 & 3.3 & 77.3 & 19.4 &  \\ 
32.08 & 0.0 & 86.1 & 13.9 &  \\ 
34.20 & 0.0 & 86.8 & 13.2 &  \\ 
35.93 & 13.8 & 70.7 & 15.4 &  \\ \hline
\multicolumn{5}{c}{(b) $J=2^{+}~T=0$ States} \\ 
{$E_{exc}(MeV)$} & $0~\hbar \omega $ & $2~\hbar \omega $ & $4~\hbar \omega $
& $B(E2)_{0_1^{+}\rightarrow 2_i^{+}}~(e^2fm^4)$ \\ \hline
3.04 & 66.3 & 23.8 & 9.9 & 65.3 \\ 
11.37 & 83.4 & 10.9 & 5.7 & 0.0 \\ 
13.59 & 86.2 & 8.9 & 4.9 & 0.0 \\ 
15.88 & 94.4 & 2.1 & 3.5 & 0.0 \\ 
15.95 & 87.5 & 8.3 & 4.2 & 0.0 \\ 
17.86 & 94.3 & 2.5 & 3.2 & 0.0 \\ 
19.39 & 94.9 & 2.1 & 3.0 & 0.0 \\ 
27.15 & 28.5 & 51.4 & 20.2 & 15.7 \\ 
30.22 & 0.0 & 79.3 & 20.7 & 0.0 \\ 
31.71 & 1.0 & 80.1 & 18.9 & 1.6 \\ 
32.09 & 0.0 & 86.1 & 13.9 & 0.0 \\ 
33.87 & 0.1 & 83.3 & 16.6 & 0.0 \\ 
34.20 & 0.0 & 86.8 & 13.2 & 0.0 \\ 
35.71 & 10.7 & 75.0 & 14.3 & 0.0 \\ 
&  &  &  & 
\end{tabular}
\end{center}

\end{table}

\begin{table}

{\bf Table II. }Same as Table I but for the interaction $-\chi Q\cdot
Q+xV_{s.o.}$ with $\chi =0.3365~MeV/fm^4$ and $x=1$.

\begin{center}
\begin{tabular}{ccccc}
\multicolumn{5}{c}{(a) $J=0^{+}~T=0$ States} \\ 
{$E_{exc}(MeV)$} & $0~\hbar \omega $ & $2~\hbar \omega $ & $4~\hbar \omega $
&  \\ \hline
0.0 & 65.1 & 24.0 & 10.9 &  \\ 
12.8 & 83.6 & 10.3 & 6.1 &  \\ 
16.4 & 89.7 & 6.0 & 4.3 &  \\ 
21.9 & 91.7 & 4.6 & 3.7 &  \\ 
26.4 & 69.3 & 21.3 & 9.4 &  \\ 
26.5 & 40.7 & 44.0 & 15.3 &  \\ 
29.9 & 3.4 & 77.4 & 19.2 &  \\ 
32.1 & 0.0 & 86.6 & 13.4 &  \\ 
37.3 & 0.0 & 85.6 & 14.3 &  \\ 
38.4 & 18.2 & 66.2 & 15.6 &  \\ \hline
\multicolumn{5}{c}{(b) $J=2^{+}~T=0$ States} \\ 
{$E_{exc}(MeV)$} & $0~\hbar \omega $ & $2~\hbar \omega $ & $4~\hbar \omega $
& $B(E2)_{0_1^{+}\rightarrow 2_i^{+}}~(e^2fm^4)$ \\ \hline
3.1 & 66.7 & 23.3 & 10.1 & 63.4 \\ 
10.2 & 85.8 & 8.8 & 5.4 & 0.4 \\ 
13.2 & 88.2 & 7.2 & 4.6 & 0.9 \\ 
16.2 & 91.9 & 4.2 & 3.9 & 0.0 \\ 
17.7 & 86.4 & 8.9 & 4.7 & 0.2 \\ 
19.6 & 88.3 & 7.4 & 4.3 & 0.0 \\ 
21.6 & 84.8 & 10.3 & 4.9 & 0.1 \\ 
22.2 & 91.0 & 5.1 & 3.8 & 0.0 \\ 
27.5 & 27.8 & 53.1 & 19.1 & 14.5 \\ 
30.9 & 0.9 & 78.0 & 21.0 & 0.0 \\ 
31.9 & 1.1 & 80.2 & 18.7 & 1.6 \\ 
32.4 & 0.0 & 86.2 & 13.8 & 0.0 \\ 
34.3 & 0.2 & 85.7 & 14.0 & 0.0 \\ 
34.6 & 1.2 & 83.8 & 15.1 & 0.1 \\ 
35.2 & 11.4 & 74.0 & 14.6 & 0.1 \\ 
&  &  &  & 
\end{tabular}
\end{center}

\end{table}

\begin{table}

\begin{center}
{\bf Table III. }Same as Table I but for the realistic ($x,y$) interaction
with $x=1$ and $y=1$. 
\begin{tabular}{ccccc}
\multicolumn{5}{c}{(a) $J=0^{+}~T=0$ States} \\ 
{$E_{exc}(MeV)$} & $0~\hbar \omega $ & $2~\hbar \omega $ & $4~\hbar \omega $
&  \\ \hline
0.0 & 62.2 & 26.2 & 11.6 &  \\ 
22.8 & 66.5 & 23.6 & 9.9 &  \\ 
28.7 & 6.5 & 71.0 & 22.5 &  \\ 
30.3 & 66.5 & 23.0 & 10.5 &  \\ 
35.3 & 67.5 & 22.4 & 10.1 &  \\ 
39.4 & 7.3 & 73.4 & 19.3 &  \\ 
43.5 & 56.3 & 31.4 & 12.3 &  \\ 
47.6 & 8.8 & 70.5 & 20.7 &  \\ 
49.5 & 2.3 & 76.7 & 21.6 &  \\ 
50.1 & 3.3 & 75.7 & 21.0 &  \\ \hline
\multicolumn{5}{c}{(b) $J=2^{+}~T=0$ States} \\ 
{$E_{exc}(MeV)$} & $0~\hbar \omega $ & $2~\hbar \omega $ & $4~\hbar \omega $
& $B(E2)_{0_1^{+}\rightarrow 2_i^{+}}~(e^2fm^4)$ \\ 
5.4 & 62.2 & 26.6 & 11.1 & 31.1 \\ 
19.5 & 70.0 & 20.4 & 9.6 & 0.0 \\ 
21.5 & 69.5 & 20.2 & 10.3 & 0.1 \\ 
26.2 & 69.7 & 20.5 & 9.8 & 0.4 \\ 
30.4 & 70.2 & 20.9 & 8.9 & 0.0 \\ 
31.0 & 56.7 & 30.9 & 12.6 & 1.7 \\ 
33.7 & 13.5 & 65.7 & 20.8 & 3.7 \\ 
35.1 & 71.3 & 19.7 & 9.0 & 0.0 \\ 
38.2 & 67.7 & 22.4 & 9.8 & 0.0 \\ 
41.6 & 9.0 & 68.8 & 22.2 & 1.3 \\ 
45.0 & 1.0 & 79.7 & 19.3 & 0.1 \\ 
45.9 & 2.9 & 77.9 & 19.2 & 2.4 \\ 
46.3 & 3.2 & 76.7 & 20.1 & 1.3 \\ 
47.3 & 0.3 & 79.5 & 20.2 & 0.0 \\ 
48.4 & 1.5 & 79.8 & 18.6 & 0.0 \\ 
&  &  &  & 
\end{tabular}
\end{center}

\end{table}

\begin{table}

{\bf Table IV. }$J=0^{+}$ and $2^{+}$ states in $^{10}Be$ for the
interaction $-\chi Q\cdot Q$ with $\chi =0.3615~MeV/fm^4$ with up to $2\hbar
\omega $ excitations allowed. The percentage of $0\hbar \omega $ and $2\hbar
\omega $ occupancies are given, as well as the $B(E2)(0_1^{+}\rightarrow
2_i^{+}$).

\begin{center}
\begin{tabular}{cccc}
& \multicolumn{2}{c}{(a) $J=0^{+}~T=1$ States} &  \\ 
{$E_{exc}(MeV)$} & $0~\hbar \omega $ & $2~\hbar \omega $ &  \\ \hline
0.0 & 81.8 & 18.2 &  \\ 
3.7 & 81.0 & 19.0 &  \\ 
7.3 & 93.6 & 6.4 &  \\ 
9.7 & 0.0 & 100.0 &  \\ 
12.1 & 92.9 & 7.1 &  \\ 
12.1 & 92.9 & 7.1 &  \\ 
13.9 & 93.1 & 6.9 &  \\ 
17.7 & 98.9 & 1.1 &  \\ 
22.1 & 0.0 & 100.0 &  \\ 
22.9 & 0.0 & 100.0 &  \\ \hline
& \multicolumn{2}{c}{(b) $J=2^{+}~T=1$ States} &  \\ 
{$E_{exc}(MeV)$} & $0~\hbar \omega $ & $2~\hbar \omega $ & 
$B(E2)_{0_1^{+}\rightarrow 2_i^{+}}~(e^2fm^4)$ \\ \hline
2.2 & 81.3 & 18.7 & 5.0 \\ 
3.4 & 83.4 & 16.6 & 47.2 \\ 
3.7 & 81.0 & 19.0 & 0.0 \\ 
7.3 & 93.6 & 6.4 & 0.0 \\ 
9.2 & 82.9 & 17.1 & 0.0 \\ 
10.9 & 91.9 & 8.1 & 0.0 \\ 
11.9 & 0.0 & 100.0 & 0.0 \\ 
12.1 & 92.9 & 7.1 & 0.0 \\ 
12.1 & 92.9 & 7.1 & 0.0 \\ 
12.1 & 92.9 & 7.1 & 0.0 \\ 
13.9 & 93.1 & 6.9 & 0.2 \\ 
13.9 & 93.1 & 6.9 & 0.0 \\ 
13.9 & 93.1 & 6.9 & 0.0 \\ 
17.7 & 98.9 & 1.1 & 0.0 \\ 
22.1 & 0.0 & 100.0 & 0.0 \\ 
&  &  & 
\end{tabular}
\end{center}

\end{table}

\begin{table}

{\bf Table V. }Same as Table IV but for the interaction $-\chi Q\cdot
Q+xV_{s.o.}$ with $\chi =0.3615~MeV/fm^4$ and $x=1$.

\begin{center}
\begin{tabular}{cccc}
& \multicolumn{2}{c}{(a) $J=0^{+}~T=1$ States} &  \\ 
{$E_{exc}(MeV)$} & $0~\hbar \omega $ & $2~\hbar \omega $ &  \\ \hline
0.0 & 85.6 & 14.4 &  \\ 
8.0 & 80.8 & 19.2 &  \\ 
9.6 & 92.0 & 8.0 &  \\ 
11.4 & 0.0 & 100.0 &  \\ 
12.1 & 91.5 & 8.5 &  \\ 
16.4 & 90.6 & 9.4 &  \\ 
19.7 & 90.5 & 9.5 &  \\ 
23.1 & 88.7 & 11.3 &  \\ 
24.0 & 0.0 & 100.0 &  \\ 
26.1 & 0.0 & 100.0 &  \\ \hline
& \multicolumn{2}{c}{(b) $J=2^{+}~T=1$ States} &  \\ 
{$E_{exc}(MeV)$} & $0~\hbar \omega $ & $2~\hbar \omega $ & 
$B(E2)_{0_1^{+}\rightarrow 2_i^{+}}~(e^2fm^4)$ \\ \hline
3.0 & 85.5 & 14.5 & 40.1 \\ 
4.6 & 83.7 & 16.3 & 3.4 \\ 
6.8 & 90.8 & 9.2 & 0.3 \\ 
7.8 & 83.5 & 16.5 & 3.7 \\ 
11.8 & 84.8 & 15.2 & 0.1 \\ 
13.0 & 91.2 & 8.8 & 0.1 \\ 
13.8 & 0.0 & 100.0 & 0.0 \\ 
14.1 & 90.9 & 9.1 & 0.0 \\ 
14.8 & 90.9 & 9.1 & 0.0 \\ 
15.5 & 90.3 & 9.7 & 0.0 \\ 
17.2 & 90.0 & 10.0 & 0.1 \\ 
17.2 & 88.0 & 12.0 & 0.0 \\ 
18.2 & 90.3 & 9.7 & 0.1 \\ 
21.2 & 89.0 & 11.0 & 0.0 \\ 
23.0 & 52.8 & 47.3 & 0.0 \\ 
&  &  & 
\end{tabular}
\end{center}

\end{table}

\begin{table}

\begin{center}
{\bf Table VI. }Same as Table IV but for the realistic ($x,y$) interaction
with $x=1$ and $y=1$. 
\begin{tabular}{cccc}
& \multicolumn{2}{c}{(a) $J=0^{+}~T=1$ States} &  \\ 
{$E_{exc}(MeV)$} & $0~\hbar \omega $ & $2~\hbar \omega $ &  \\ \hline
0.0 & 73.3 & 26.7 &  \\ 
8.7 & 74.4 & 25.6 &  \\ 
12.0 & 74.7 & 25.3 &  \\ 
21.1 & 76.5 & 23.5 &  \\ 
23.7 & 77.5 & 22.5 &  \\ 
31.0 & 49.3 & 50.7 &  \\ 
31.5 & 25.4 & 74.6 &  \\ 
34.5 & 5.8 & 94.2 &  \\ 
37.6 & 0.6 & 99.4 &  \\ 
39.7 & 74.1 & 25.9 &  \\ \hline
& \multicolumn{2}{c}{(b) $J=2^{+}~T=1$ States} &  \\ 
{$E_{exc}(MeV)$} & $0~\hbar \omega $ & $2~\hbar \omega $ & 
$B(E2)_{0_1^{+}\rightarrow 2_i^{+}}~(e^2fm^4)$ \\ \hline
4.6 & 73.5 & 26.5 & 19.7 \\ 
5.2 & 73.9 & 26.1 & 3.2 \\ 
9.2 & 73.7 & 26.3 & 1.5 \\ 
10.1 & 75.8 & 24.2 & 0.0 \\ 
17.4 & 74.5 & 25.5 & 0.0 \\ 
19.7 & 75.7 & 24.3 & 0.1 \\ 
20.2 & 77.0 & 23.0 & 0.0 \\ 
22.1 & 76.9 & 23.1 & 0.2 \\ 
22.9 & 77.1 & 22.9 & 0.0 \\ 
23.7 & 77.2 & 22.8 & 0.0 \\ 
27.2 & 76.8 & 23.2 & 0.0 \\ 
29.0 & 76.9 & 23.1 & 0.2 \\ 
32.5 & 76.9 & 23.1 & 0.0 \\ 
33.4 & 0.3 & 99.7 & 0.0 \\ 
35.5 & 71.7 & 28.3 & 0.2 \\ 
&  &  & 
\end{tabular}
\end{center}

\end{table}

\begin{table}

{\bf Table VII. }Excitation energies (in $MeV$) of the first $J=0^{+}$ and
$2^{+}$ intruder states in $^8Be$ and $^{10}Be$.

\begin{center}
\begin{tabular}{cccc}
& $Q\cdot Q$ & $Q\cdot Q~+~xV_{s.o.}$ & ($x,y$)=(1,1) \\ \hline
\multicolumn{4}{c}{$^8Be~~J=0^{+}~T=0$} \\ 
$2\hbar \omega $ & 32.1 & 30.1 & 33.8 \\ 
$4\hbar \omega $ & 26.5 & 26.5 & 28.7 \\ \hline
\multicolumn{4}{c}{$^8Be~~J=2^{+}~T=0$} \\ 
$2\hbar \omega $ & 31.5 & 30.9 & 36.6 \\ 
$4\hbar \omega $ & 27.5 & 27.5 & 33.7 \\ \hline
\multicolumn{4}{c}{$^{10}Be~~J=0^{+}~T=1$} \\ 
$2\hbar \omega $ & 9.7 & 11.4 & 31.0 \\ \hline
\multicolumn{4}{c}{$^{10}Be~~J=2^{+}~T=1$} \\ 
$2\hbar \omega $ & 11.9 & 13.8 & 33.6 \\ 
&  &  & 
\end{tabular}
\end{center}

\end{table}

\begin{table}

{\bf Table VIII. }Excitation energies of the first $J=0^{+}$ and $2^{+}$
intruder states in $^8Be$, $^{10}Be$, and $^{12}C$ in the deformed
oscillator model.

\begin{center}
\begin{tabular}{clcccccccccc}
&  & $\Sigma _x,\Sigma _y,\Sigma _z$ & $\frac{\omega _x}{\omega _0}, 
\frac{\omega _y}{\omega _0},\frac{\omega _z}{\omega _0}$ & $E_{int}\,$ & 
$\left\langle J_x^2\right\rangle ,\left\langle J_y^2\right\rangle
,\left\langle J_z^2\right\rangle $ & ${\cal I}_x,{\cal I}_y,{\cal I}_z 
{\cal \,}$ & $\Delta E_R$ & $E_{J=0}^{*}\,$ & $\hbar \omega _0$ & 
$E_{J=0}^{*}\,$ & $E_{J=2}^{*}\,$ \\ 
&  &  &  & $\left[ \hbar \omega _0\right] $ &  & $\left[ (\hbar \omega
_0)^{-1}\right] $ & $\left[ \hbar \omega _0\right] $ & $\left[ \hbar
\omega_0\right] $ & [MeV] & [MeV] & [MeV] \\ \hline
$^8Be$ & $0p-0h$ & 4,4,8 & 1.26,1.26,0.63 & 15.12 & 6,6,0 & 15.9,15.9,0 & 
0.38 &  & 16.25 &  & 3.07 \\ 
& $2p-2h$ & 4,4,10 & 1.36,1.36,0.54 & 16.29 & 10.5,10.5,0 & 21.4,21.4,0 & 
0.49 & 1.06 & 16.25 & 17.23 & 19.51 \\ 
& $4p-4h$ & 4,4,12 & 1.44,1.44,0.48 & 17.31 & 16,16,0 & 27.7,27.7,0 & 0.58 & 
1.99 & 16.25 & 32.34 & 34.10 \\ \hline
$^{10}Be$ & $\left( 0p-0h\right) _{triaxial}$ & 7,5,9 & 0.97,1.36,0.76 & 
20.41 & 5.6,2.3,2.4 & 15.6,19.2,10.8 & 0.70 &  & 15.50 &  & 2.70 \\ 
& $2p-2h$ & 5,5,13 & 1.38,1.38,0.53 & 20.63 & 14.4,14.4,0 & 28.2,28.2,0 & 
0.51 & 0.41 & 15.50 & 6.36 & 8.01 \\ 
& $\left( 0p-0h\right) _{axial}$ & 6,6,9 & 1.14,1.14,0.76 & 20.61 & 
3.75,3.75,0 & 17.0,17.0,0 & 0.22 & 0.67 & 15.50 & 10.39 & 13.12 \\ \hline
$^{12}C$ & $0p-0h$ & 10,6,10 & 0.84,1.41,0.84 & 25.30 & 5.3,0,5.3 & 
16.1,0,16.1 & 0.33 &  & 14.89 &  & 2.77 \\ 
& $4p-4h$ & 6,6,18 & 1.44,1.44,0.48 & 25.96 & 21,21,0 & 38.5,38.5,0 & 0.55 & 
0.44 & 14.89 & 6.55 & 7.71
\end{tabular}
\end{center}

\end{table}

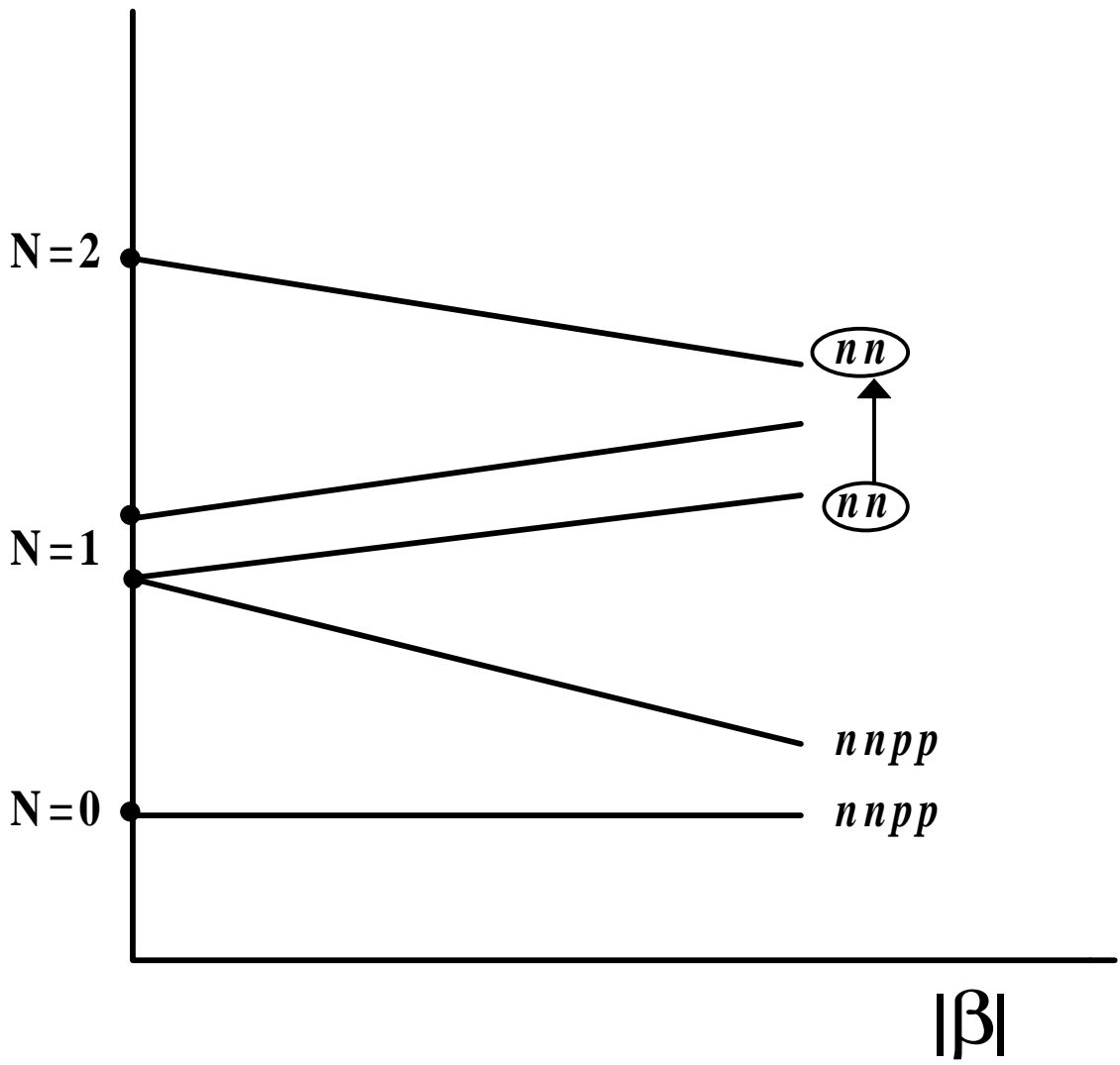
\end{document}